\RequirePackage{lineno}
\documentclass[aps,prl,twocolumn,superscriptaddress]{revtex4}
\usepackage{mathrsfs,amssymb,graphics,subfigure,threeparttable}
\usepackage{grap hicx}
\usepackage{amssymb}
\usepackage{enumerate}
\usepackage{amsmath}
\usepackage{fancyhdr}
\usepackage{bm}
\usepackage[squaren]{SIunits}

\begin{document}
%\setpagewiselinenumbers
%\modulolinenumbers[1] 
%\linenumbers  
%\begin{frontmatter} 
%\preprint{}


\title{A near-ideal dechirper for plasma-based electron and positron acceleration using a hollow channel plasma}


\author{Y. P. Wu}
\affiliation{Tsinghua University, Beijing 100084, China}
\author{J. F. Hua}
\email[]{jfhua@tsinghua.edu.cn}
\affiliation{Tsinghua University, Beijing 100084, China}
\author{C. H. Pai}
\affiliation{Tsinghua University, Beijing 100084, China}
\author{W. An}
\affiliation{Beijing Normal University, Beijing 100875, China}
\author{Z. Zhou}
\affiliation{Tsinghua University, Beijing 100084, China}
\author{J. Zhang}
\affiliation{Tsinghua University, Beijing 100084, China}
\author{S. Liu}
\affiliation{Tsinghua University, Beijing 100084, China}
\author{B. Peng}
\affiliation{Tsinghua University, Beijing 100084, China}
\author{Y. Fang}
\affiliation{Tsinghua University, Beijing 100084, China}
\author{S. Y. Zhou}
\affiliation{Tsinghua University, Beijing 100084, China}
\author{X. L. Xu}
\affiliation{SLAC National Accelerator Laboratory, Standford, California 94309, USA}
\author{C. J. Zhang}
\affiliation{University of California Los Angeles, Los Angeles, California 90095, USA}
\author{F. Li}
\affiliation{University of California Los Angeles, Los Angeles, California 90095, USA}
\author{Z. Nie}
\affiliation{University of California Los Angeles, Los Angeles, California 90095, USA}
\author{W. Lu}
\email[]{weilu@tsinghua.edu.cn}
\affiliation{Tsinghua University, Beijing 100084, China}
\author{W. B. Mori}
\affiliation{University of California Los Angeles, Los Angeles, California 90095, USA}
\author{C. Joshi}
\affiliation{University of California Los Angeles, Los Angeles, California 90095, USA}


\date{\today}

\begin{abstract}
Plasma-based electron/positron wakefield acceleration has made great strides in the past decade. Currently, one major challenge for its applications to coherent light sources and colliders is the relatively large energy spread (few percent level) of the accelerated beams. This energy spread is usually dominated by a longitudinally correlated energy chirp induced by the wakefield. It is highly desirable to have a systematic method for reducing this spread down to $\lesssim 0.1\%$, and at the same time maintaining the beam emittance. 
In this paper, a near-ideal dechirper based on a low-density hollow channel plasma is proposed and tested through large-scale three-dimensional particle-in-cell simulations.
The theoretical analyses and simulations confirm that the positively correlated energy spread (chirp) of the electron/positron beam from a plasma accelerator can be significantly reduced by its self-wake in the hollow plasma channel from a few percent down to $\lesssim 0.1\%$ without noticeable beam emittance growth induced by the transverse fields. 
Such near-ideal dechirper may significantly improve the beam quality of plasma-based accelerators, paving the way for their applications to future compact free electron lasers and colliders.
\end{abstract}

%\keywords{}

\maketitle

\section{I. Introduction}
A plasma wake driven by an intense laser pulse or a charged particle beam can be utilized to accelerate electrons and positrons at extremely large accelerating fields of $\sim$ 100 GV/m, orders of magnitude larger than those in state-of-the-art radio-frequency microwave-based accelerators \cite{PhysRevLett.43.267, Chen_PRL1985}.
In the past decade, the field of plasma-based wakefield acceleration has reached several key milestones, such as multi-GeV electron acceleration in laser driven wakes \cite{Nature_2004_1, Nature_2004_2, Nature_2004_3, leemans2006gev, Nature_Photonics_2008_gev, wang2013quasi, PRL_3GeV, leemans_4gev, BELLA_8GeV}, and high-energy, high-efficiency electron/positron acceleration in beam driven wakes \cite{blumenfeld2007energy, Nature_high_efficiency, Nature_positron, NC_positron_driven_hollow_channel_PWFA}, showing its potentials for compact X-ray free electron lasers (X-FELs) and linear colliders for particle physics.
For these challenging applications, beams with both low emittance and low energy spread are required.
For beam emittance, currently sub-$\mu$m rad emittance has been measured \cite{emittance_measurement_1, emittance_measurement_2} in laser wakefield acceleration experiments, and studies on controlled injection schemes also show the potential of generating beams with much lower emittance \cite{Hidding_injection, LifeiInjection, XinluTwoColorInjection, Down-ramp_Injection}.
For the energy spread, however, current experiments can only produce beams with percent level energy spread, and this is about one order of magnitude larger than what is required for X-FELs and colliders.
Therefore, it is critical to find systematic solutions for reducing the beam energy spread down to 0.1 percent or lower.

Before one can invent ways to reduce the energy spread, it is imperative to understand where the energy spread comes from. In order to obtain high gradients, the plasma accelerator must be operated in a high-density plasma which in turn means that the wavelength of the accelerating structure is microscopic (tens to hundreds of micrometers).
Due to the very short wavelength of this structure, the acceleration phase interval occupied by the short electron/positron beams is much larger than that in a traditional accelerator. This can lead to a significant energy chirp in the beam unless ideal beam loading can be achieved \cite{BeamLoading1987, PhysRevLett.101.145002}. 
This chirp is typically much larger than the intrinsic slice energy spread of the beam. For example, in many recently proposed injection schemes, the intrinsic energy spread of the beam slices could be reduced down to $\sim1$ MeV or even  tens of keV level \cite{LifeiInjection, XinluTwoColorInjection, Down-ramp_Injection, Lifei_NIT_injection, CollidingIonizationInjection}, therefore, for beam energies more than a few hundred MeV, the energy chirp induced by the acceleration phase variance becomes the dominant part of the total energy spread.

\begin{figure}[tp]
\includegraphics[height=0.26\textwidth]{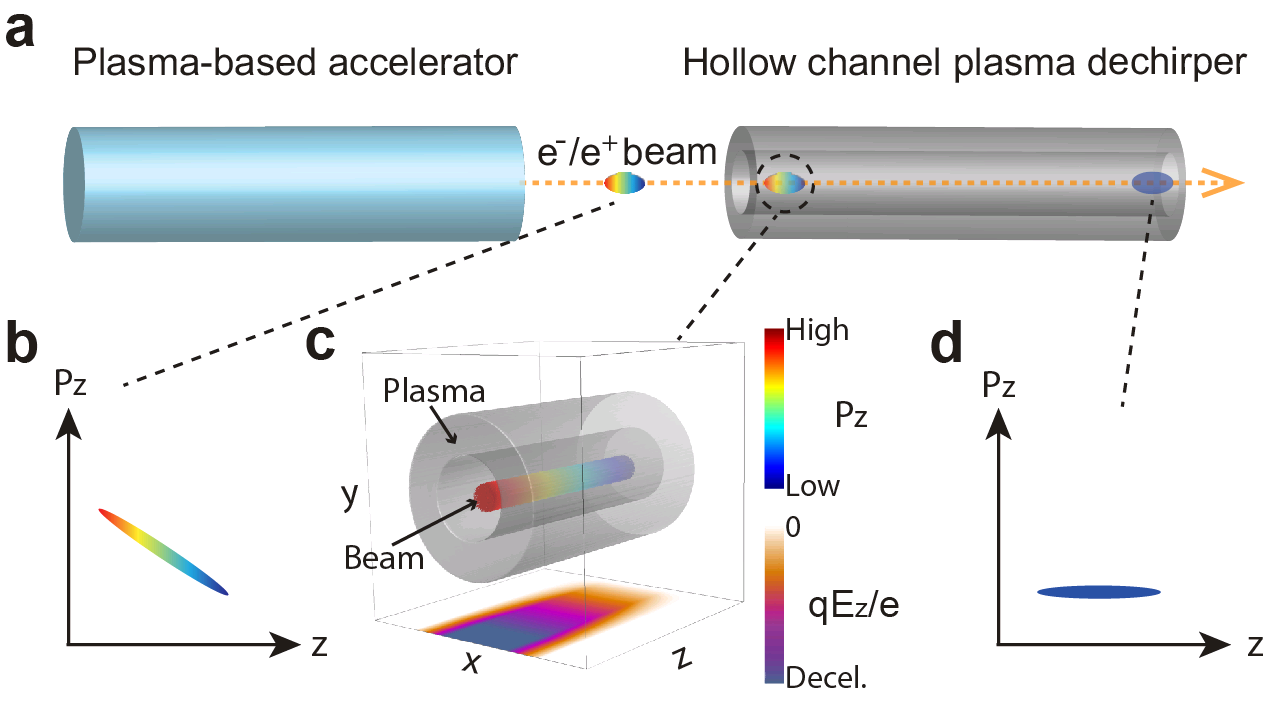}
\caption{\label{fig1}
Schematic diagram for a near-ideal hollow channel plasma dechirper. (a) A positively chirped electron/positron beam is generated in a plasma accelerator at first, and then sent through a hollow channel plasma to dechirp itself. (b)The initial beam longitudinal phasespace ($z$ versus $p_z$). (c) The longitudinal wakefield $E_z$ excited by the beam in the hollow channel plasma.
The projection in the $x-z$ plane shows the slice of $qE_z/e$ at the mid-plane in the $y$ direction, where $q$ is the particle charge ($-e$ for electron and $+e$ for positron).
(d) The final beam longitudinal phasespace ($z$ versus $p_z$).}
\end{figure}

To reduce the relatively large energy chirp in plasma accelerators, energy dechirpers based on tenuous uniform plasmas have recently been proposed and demonstrated \cite{dechirper_IPAC_paper, THU_dechirper, FlashForward_dechirper, INFN_dechirper}, and phase space measurements 
 clearly show the energy spread reduction effectiveness of such dechirper \cite{THU_dechirper}.
However, this approach has a limitation for maintaining emittances well below $\sim$1 $\mu$m rad due to the longitudinal-position-dependent transverse wakefield focusing, and this effect is especially severe for positron beams.
In addition, dechirping in a uniform plasma also induces increasing slice energy spread due to the nonuniform longitudinal wakefield in the transverse dimensions.
These two factors impose restrictions for its application to X-FELs and linear colliders.
Ideally, a dechirper that can both remove the energy chirp and maintain the emittance is preferred.
In this paper, we propose to use a low-density hollow channel plasma to serve as such a near-ideal dechirper, as shown in the schematic diagram in Fig. \ref{fig1}(a).

In this scheme, a beam with a nearly linear positive energy chirp [Fig. 1(b), the beam energy increases quasi-linearly from head to tail, which is normal for an underloaded wake in a plasma accelerator \cite{LifeiInjection, NC_chirp_compensation}] is sent through a separate low-density hollow channel plasma section to excite a nearly linear plasma wake [Fig. 1(c)].
If the wake wavelength is much longer than the bunch length, the beam will totally stay in a decelerating phase of the wake with a negative slope (the tail of the beam experiences greater energy loss gradient than the head) inside the channel. 
Such a decelerating wake can effectively reduce the beam positive chirp during the propagation [Fig. 1(d)].
The total dechirping effects can be easily tuned by changing the density, length and geometry (i.e., inner and outer radii) of the plasma channel. 
If the parameters are properly designed, an energy spread reduction down to $0.1\%$ level or even lower is indeed possible. At the same time, the transverse focusing fields inside the channel will be zero or negligibly small if the beam is launched on or very close to the axis, and this will help to preserve the beam emittance
\cite{high_quality_and_efficiency_hollow_channel_plasma_accelerator}.
We also note that this scheme works equally well for electron and positron beams, a unique feature crucial for the application to electron/positron colliders.

In the following sections, detailed theoretical analyses and three-dimensional (3D) particle-in-cell (PIC) simulations will be systematically presented to show the effectiveness of the above scheme on energy spread reduction down to $0.1\%$ level and emittance preservation.

\section{II. Concept and PIC simulation illustration}
To illustrate the effectiveness of the hollow channel plasma dechirper, we show here one example through 3D PIC simulations using the code QuickPIC \cite{QuickPIC_1, QuickPIC_2, QuickPIC_OpenSource}. In this example, a 4 GeV electron/positron beam with a $1\%$ (RMS) linear positive energy chirp and an intrinsic slice energy spread of 0.4 MeV (RMS) is sent on-axis through a $\sim$ 55 cm long hollow channel plasma.
The beam has 1 nC charge with a high peak current $I_b=10$ kA (near flat-top current profile) and low normalized emittance $\epsilon_{nx,y}=0.2$ $\mu$m rad. 
As we will see later energy chirps for other current profiles can also be removed but not as completely as for a flat-top pulse.
These parameters are chosen to be comparable to the required high-quality beam parameters in a future linear collider design \cite{ILC_TDR_book, LWFA_collider}.
The hollow channel plasma has an inner radius $a=300\ \mu$m, outer radius $b=500\ \mu$m,
and electron density $n_p=5\times10^{15}$ cm$^{-3}$ within the annular plasma ring.
By the end of the simulations, the relative energy spreads ($\delta_W$) of both the electron and the positron beams have been dramatically reduced from $1\%$ to $\sim 0.02\%$, close to the intrinsic slice energy spread ($0.01\%$).
At the same time, the emittances of the beams remain almost unchanged.
Therefore the 6D-brightness of the beam [$B_{6D}=2I_b/(\epsilon_{nx}\epsilon_{ny}\delta_W)\approx2.5\times 10^{18}{\rm Am^{-2}rad^{-2}}/0.1\%$] has been effectively enhanced by a factor of 50.

\begin{figure*}[tp]
\includegraphics[height=0.5\textwidth]{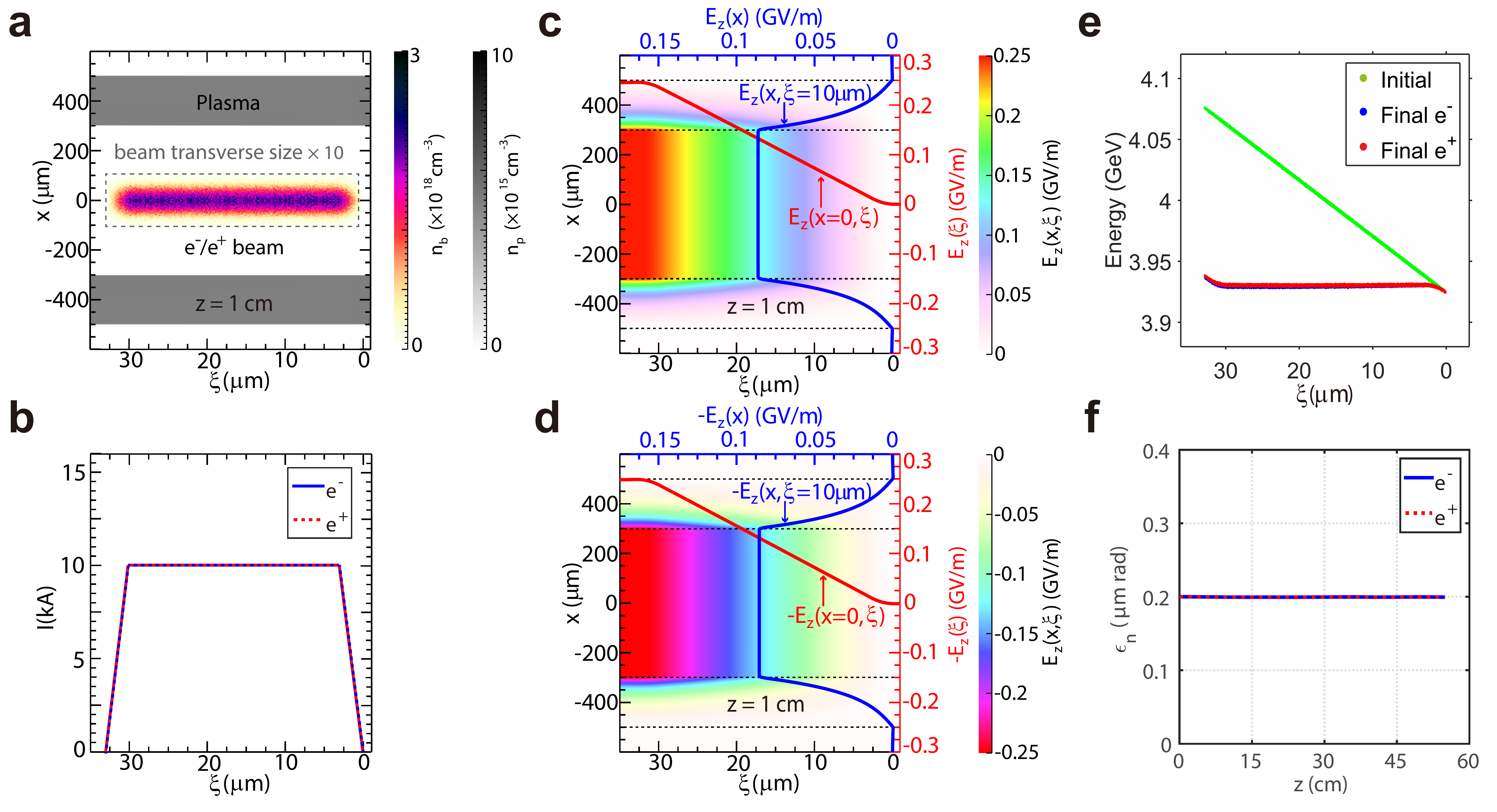}
\caption{\label{fig2}
3D PIC simulations of the dechirping process in a hollow channel plasma.
(a) The densities of the plasma channel ($n_p$) and the e$^{-}$/e$^{+}$ beam ($n_b$) in the $x-\xi$ plane 
when the propagation distance $z=1$ cm. Here $\xi=ct-z$ represents the longitudinal position relative to the beam. 
(b) The e$^{-}$/e$^{+}$ beam current profiles (a $90$ fs long plateau and two $10$ fs long ramps at each sides). (c) and (d) show the $E_z$ field excited by the e$^{-}$ and e$^{+}$ bunches, respectively, also at $z=1$ cm. The black dotted lines represent the inner and outer radii of the plasma channel. Lineouts of the on-axis $E_z$ ($x=0\ \mu$m, $\xi$) and the radial variation in $E_z$ ($x$, $\xi=10\ \mu$m) are separately shown with red and blue lines.
(e) The longitudinal phase spaces of e$^{-}$/e$^{+}$ beams before and after the dechirper.
(f) The evolutions of e$^{-}$/e$^{+}$ beam normalized emittances during the dechirping process.
}
\end{figure*}

The details of the simulations are presented in Fig. \ref{fig2}. The initial densities of the plasma channel and the e$^-$/e$^+$ beam are plotted in Fig. \ref{fig2} (a). The beams have a transverse Gaussian profile with $\sigma_{x,y}=4.0$ $\mu$m, and a near flat-top longitudinal current profile [Fig. \ref{fig2} (b)].
The longitudinal wakefield $E_z$s in the hollow channel plasma excited by the e$^{-}$/e$^{+}$ beams are shown in Fig. \ref{fig2} (c) and (d), respectively.
One can see that these two fields are very similar except for a change of charge sign, and the on-axis lineouts of both $E_z$ have a near linear deceleration along most of the beam length.
The transverse uniformity of $E_z$ within the channel can also be readily seen in Fig. \ref{fig2} (c) and (d).
Combining the above two features of $E_z$, both beams (e$^{-}$/e$^{+}$) can be dechirped by the annular plasma with negligible slice energy spread increase.
After a propagation of 55.3 cm (e$^{-}$)/54.6 cm (e$^{+}$), the minimum RMS energy spread of 0.76 MeV (e$^{-}$) / 0.72 MeV( e$^{+}$) is achieved with a mean energy of 3.93 GeV.
Thus the relative energy spreads of the beams have been reduced from $1\%$ down to $0.019\%$(e$^{-}$)/ $0.018\%$ (e$^{+}$). We note that the subtle differences between the e$^{-}$ and e$^{+}$ beams come from the slightly different nonlinear plasma response \cite{thesis_the_positron_driven_hollow_channel_PWFA}.
Figure \ref{fig2} (e) shows the corresponding longitudinal phase spaces of e$^{-}$ and e$^{+}$ bunches before and after the dechirper.
Clearly the incoming linear energy chirp has been removed except at the very front and rear of the bunch.
In Fig. \ref{fig2} (f), the evolutions of e$^{-}$/e$^{+}$ beam normalized emittances over the whole propagation are plotted, and it is evident that the emittances are well conserved.

\begin{figure*}[tp]
\includegraphics[height=0.42\textwidth]{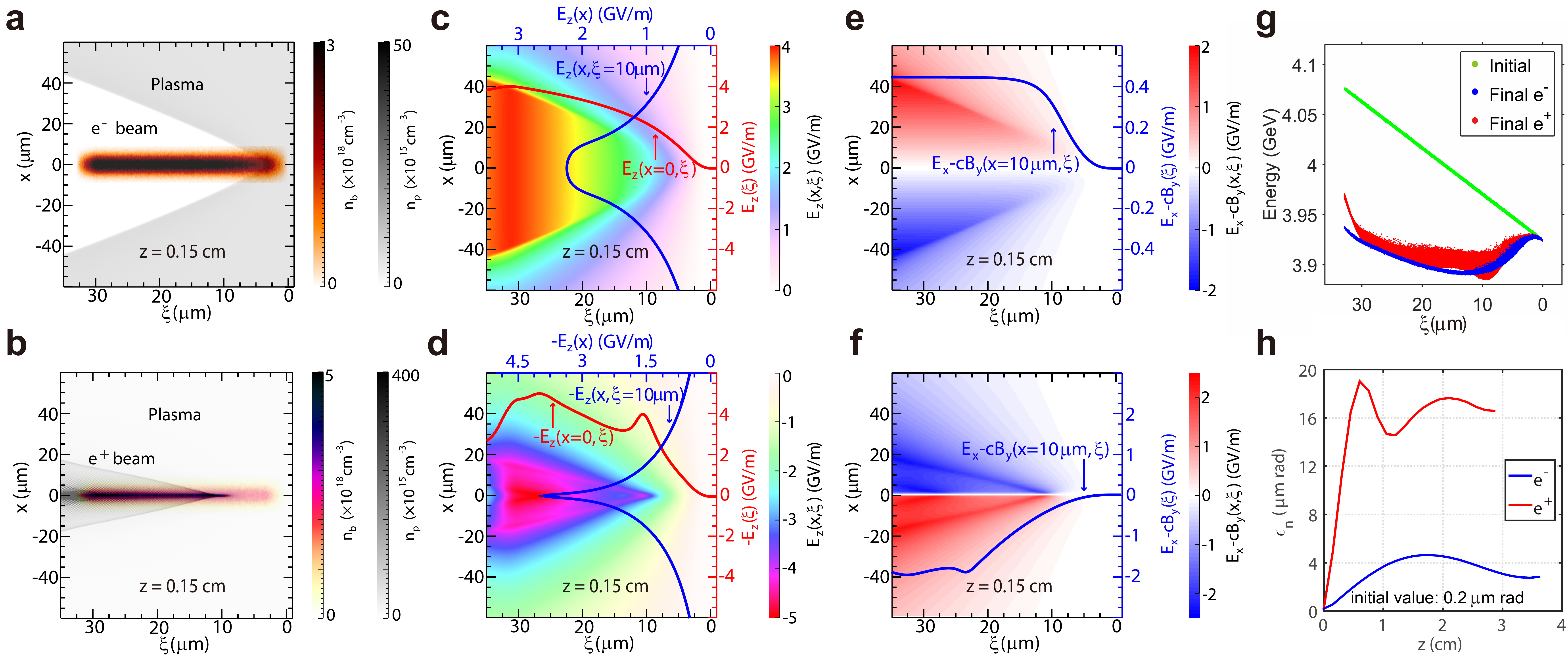}
\caption{\label{fig3}
3D PIC simulations of the dechirping process in a uniform plasma.
(a) and (b) show the densities of the plasma ($n_p$) and the beam ($n_b$) in the $x-\xi$ plane for e$^{-}$ and e$^{+}$ cases, respectively, at $z=0.15$ cm. 
(c) and (d) show the $E_z$ field excited by the e$^{-}$ and e$^{+}$ bunches, respectively, also at $z=0.15$ cm. Lineouts of the on-axis $E_z$ ($x=0\ \mu$m, $\xi$) and the radial variation in $E_z$ ($x$, $\xi=10\ \mu$m) are separately shown with red and blue lines.
(e) and (f) show the transverse wakefield $E_x-cB_y$ excited by the e$^{-}$ and e$^{+}$ bunches, respectively, also at $z=0.15$ cm. The lineout of the off-axis $E_x-cB_y$ ($x=10\ \mu$m, $\xi$) is shown with the blue line.
(g) The longitudinal phase spaces of e$^{-}$/e$^{+}$ beam before and after the dechirper.
(h) The evolutions of e$^{-}$/e$^{+}$ beam normalized emittances during the dechirping process.
}
\end{figure*}

To illustrate the differences between the hollow channel and the recently proposed uniform plasma dechirper schemes, we also simulate the dechirping process of the same e$^{-}$/e$^{+}$ beam in a uniform plasma with the density identical to the above example. 
Fig. 3 (a)/(b) shows the density distributions of the e$^{-}$/e$^{+}$ beam and the plasma. 
Due to such high beam density ($n_b\gg n_p$), the plasma has a very different nonlinear response to the e$^{-}$ and e$^{+}$ beams \cite{LimitsOfLinearPlasmaWakefieldTheory}. The e$^{-}$ beam head blows plasma electrons away and thus its tail is located in an ion channel. In comparison, the e$^{+}$ beam sucks plasma electrons in, forming a highly dense electron sheath in the beam region. The resulting longitudinal wakefield $E_z$s excited by the e$^{-}$ and e$^{+}$ beams are shown in Fig. 3 (c) and (d), respectively, 
where the longitudinal nonlinearity and the transverse nonuniformity (especially for the e$^{+}$ case) of $E_z$ are obvious. These two features will induce nonlinear energy chirp growth and slice energy spread increase during the dechirping process. 
After a propagation of 3.61 cm (e$^{-}$)/2.86 cm (e$^{+}$), the minimum RMS energy spread of 11.58 MeV (e$^{-}$) / 12.49 MeV( e$^{+}$) is obtained, which is about 15 (e$^{-}$)/17 (e$^{+}$) times larger than that achieved in the corresponding hollow channel plasma dechirper case. The longitudinal phase spaces of e$^{-}$ and e$^{+}$ bunch before and after the dechirper are shown in Fig. 3 (g), 
where the induced nonlinear energy chirp increase and slice energy spread growth are evident.
Fig. 3 (e) and (f) show the transverse wakefield $W_\perp=E_x-cB_y$ excited by the the e$^{-}$ and e$^{+}$ bunches, respectively. 
One can see that, these two fields both depend on the longitudinal position, and thus electrons at different longitudinal positions feel different focusing strength and rotate with different velocities in the transverse phase space, leading to a large growth in beam projected emittance.
In Fig. 3 (h), the evolutions of e$^{-}$/e$^{+}$ beam normalized emittance over the whole propagation are plotted, and it is clear that the emittance has increased by a factor of 14 (e$^{-}$)/ 83 (e$^{+}$). 
Such enormous emittance growth is catastrophic for beam quality.

The above comparisons show that the hollow channel plasma dechirper scheme has a decisive advantage over the uniform plasma dechiper scheme in energy spread reduction and emittance preservation for high-brightness e$^{-}$/e$^{+}$ beams. 
Next we will analyze this concept in details with theoretical analyses and 3D PIC simulations for various beam parameters.

\section{III. Theory and PIC simulation verification} 

\begin{table*}[tp]
\begin{threeparttable}
    \caption{\label{tab1}Dechirping effects for three typical current profiles\tnote{$\ddagger$}.}
    \begin{tabular}{l|c|c|c}
        \hline\hline

         &   Flat-top & Sin$^2$ & Gaussian\\
        \hline
        $f(\xi)$ & 1 & $\sin^2(\pi\xi/L_b)$  &   $\mathrm{e}^{ -\frac{(\xi-L_b/2)^2}{2\times(L_b/6)^2}}$\tnote{$\star$}\\
       $E_z(r,\xi)$ & $-\frac{q}{e}\frac{m k_p^2 c^2}{e}A_0\frac{I_b}{I_A} \xi$ & $-\frac{q}{e}\frac{m k_p^2 c^2}{e}A_0\frac{I_b}{I_A}\times \left [\frac{\xi}{2}-\frac{L_b\sin(2\pi \xi/L_b)}{4\pi}\right ]$  &   $-\frac{q}{e}\frac{m k_p^2 c^2}{e}A_0\frac{I_b}{I_A}\times \sqrt{\frac{\pi}{72}}L_b \left [1+\mathrm{Erf}\left (\frac{\xi-L_b/2}{\sqrt{2}L_b/6} \right ) \right ]$\tnote{$\dagger$}\\
      $G$ & $G_F=0$& $G_S  \approx 0.139 $  &   $G_G  \approx 0.223$\\
       $H$ & $H_F\approx 3.464$ & $H_S  \approx 3.430$  &   $H_G\approx 3.377$\\
        \hline
        \hline
    \end{tabular}
\begin{tablenotes}
\item[$\ddagger$] The subscripts ``F", ``S" and ``G" refer to the flat-top, sin$^2$ and Gaussian current profiles, respectively.
\item[$\star$]  The RMS bunch length is assumed to be $L_b/6$.
\item[$\dagger$] $\mathrm{Erf}(\xi)=\frac{2}{\sqrt{\pi}} \sum_{n=0}^\infty\frac{(-1)^n\xi^{2n+1}}{n!(2n+1)}$ is the Gauss error function.
\end{tablenotes}
\end{threeparttable}
\end{table*}

\begin{figure*}[tp]
\includegraphics[height=0.52\textwidth]{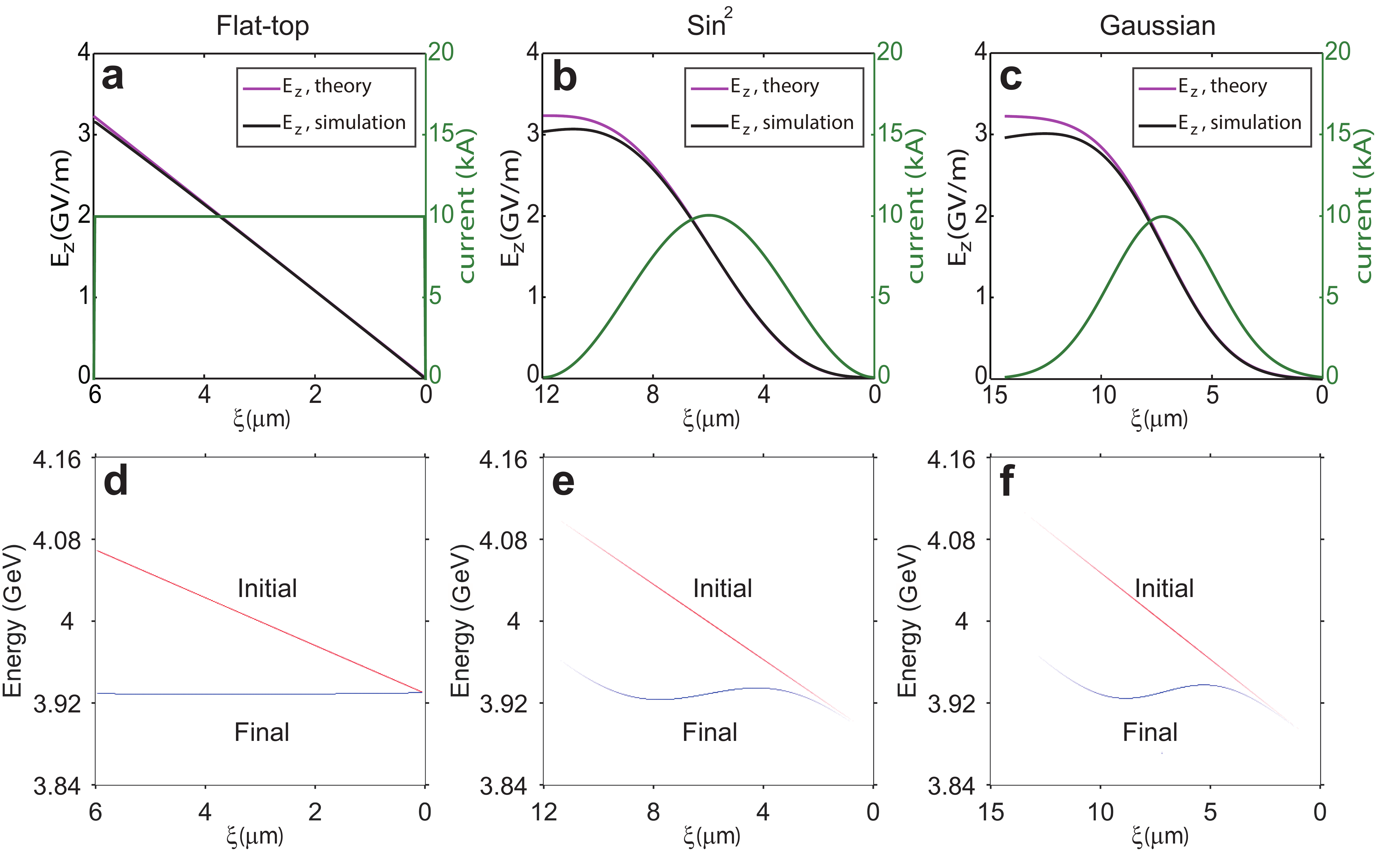}
\caption{\label{fig4}
Dechirping effects for three typical beam current profiles.
Lineouts of the calculated (in purple) and simulated (in black) $E_z$ field for flat-top (a), sin$^2$ (b) and Gaussian (c) current profiles. The corresponding beam longitudinal phasespaces before (in red) and after (in blue) the dechirper for flat-top (d), sin$^2$ (e) and Gaussian (f) current profiles.
The beam parameters of bunch charge ($Q=200$ pC), beam peak current ($I_b=10$ kA), transverse beam size ($\sigma_{x,y}=1$ $\mu$m) and normalized emittance ($\epsilon_{nx,y}=50$ nm rad) are all the same for these three cases, which give the corresponding full bunch length $L_b$ ($L_{b,F}=6$ $\mu$m for flat-top, $L_{b,S}=12$ $\mu$m for sin$^2$ and $L_{b,G}\approx 14.4$ $\mu$m for Gaussian).  
}
\end{figure*}

To quantify the effectiveness of the hollow channel plasma dechirper for various beam parameters, we have carried out a theoretical analysis based on the linear wakefield theory \cite{BeamLoading1987, PRL_hollow_plasma_channel_wakefield_accelerator}. 
In the short bunch limit where the wake wavelength is much larger than the beam bunch length, the plasma response to a narrow drive bunch in the hollow channel is relatively weak, therefore the linear plasma wakefield theory can be adopted to properly describe the wakefield structure within the beam.

Using the linear wakefield theory, the longitudinal wakefield $E_z$ in the hollow channel can be expressed as a convolution of the bunch charge distribution with a single-particle wakefuntion \cite{PRL_hollow_plasma_channel_wakefield_accelerator, NC_positron_driven_hollow_channel_PWFA, thesis_the_positron_driven_hollow_channel_PWFA}
\begin{align}\label{eq1}
E_z(r,\xi)=-\frac{q}{e}\frac{m k_p^2 c^2}{e}A_0 \int_{-\infty}^{\xi}d\xi'\cos[\Omega_0k_p(\xi-\xi')]\frac{I(\xi')}{I_A}
\end{align}
where $m$ is the electron rest mass, $k_p=\sqrt{n_pe^2/m\varepsilon_0c^2}$ is the plasma wavenumber, $\varepsilon_0$ is the vacuum permittivity, $I(\xi)$ is the beam current and $I_A\approx 17$ kA is the Alfven current. In this equation, $A_0$ and $\Omega_0$ are two quantities related to the wake amplitude and wavelength 
\begin{align}\label{eq2}
A_0 = \frac{-4B_{00}(a,b)}{k_pa[2B_{10}(a,b)-k_paB_{00}(a,b)]}
\end{align}
\begin{align}\label{eq3}
\Omega_0 = \sqrt{\frac{2B_{10}(a,b)}{2B_{10}(a,b)-k_paB_{00}(a,b)}}
\end{align}
where $B_{ij}(a,b)=(-1)^{i-j+1}I_j(k_pb)K_i(k_pa)+I_i(k_pa)K_j(k_pb)$, and $K_n$ and $I_n$ are the modified Bessel functions of order $n$.

In the short bunch limit ($k_pL_b\ll 1$), where $L_b$ is the full bunch length, 
$\cos[\Omega_0k_p(\xi-\xi')]$ reduces to $1$, therefore within the beam,
Eq. \eqref{eq1} can be simplified as
\begin{align}\label{eq4}
E_z(r,\xi)\approx -\frac{q}{e}\frac{m k_p^2 c^2}{e}A_0\frac{I_b}{I_A} \int_{0}^{\xi}d\xi'f(\xi')
\end{align}
Here $f(\xi)$ is the normalized current profile.
Equation \eqref{eq4} shows that $E_z$ is a decelerating field with a negative slope for both e$^{-}$ and e$^{+}$ beams. 
To quantify the dependence of $E_z$ on the current profile, the expressions of $E_z$ for three typical profiles (flat-top, sin$^2$ and Gaussian) are calculated and listed in Table \ref{tab1} (see details in Ref. \cite{supplemental_material}).
For the flat-top current profile, $E_z$ within the beam is exactly linear along $\xi$, which is ideal for completely removing a linear energy chirp.
For non-flat-top profiles, nonlinear energy chirps will be induced, therefore the final achievable minimum energy spread  $\Delta W_{f}$ will be a trade-off between the linear chirp reduction and the nonlinear chirp growth.

$\Delta W_{f}$ and the required channel length $L_c$ can be calculated as $\Delta W_{f}=\Delta W_{i}G$ and $L_c=\frac{\Delta W_{i}}{|qE_z(r,\xi=L_b)|}H$, where $\Delta W_{i}$ is the initial energy spread (RMS), $E_z(r,\xi=L_b)$ is the longitudinal electric field at the tail of the beam, $G$ and $H$ are two geometrical factors determined by the beam current profile $f(\xi)$. The expressions of $G$ and $H$ for these three profiles are also listed in Table \ref{tab1} (see details in Ref. \cite{supplemental_material}).

To verify the above theoretical expressions, a series of 3D PIC simulations using the code QuickPIC have been performed.
In these simulations, electron beams with different peak currents and profiles are initialized with zero slice energy spread and positive linear energy chirp [mean energy of 4 GeV and relative energy spread of $1\%$ (RMS)].
The density of the annular plasma channel is $n_p=1\times10^{17}$ cm$^{-3}$ with inner radius $a=30$ $\mu$m and outer radius $b=60$ $\mu$m.

\begin{figure*}[tp]
\includegraphics[height=0.52\textwidth]{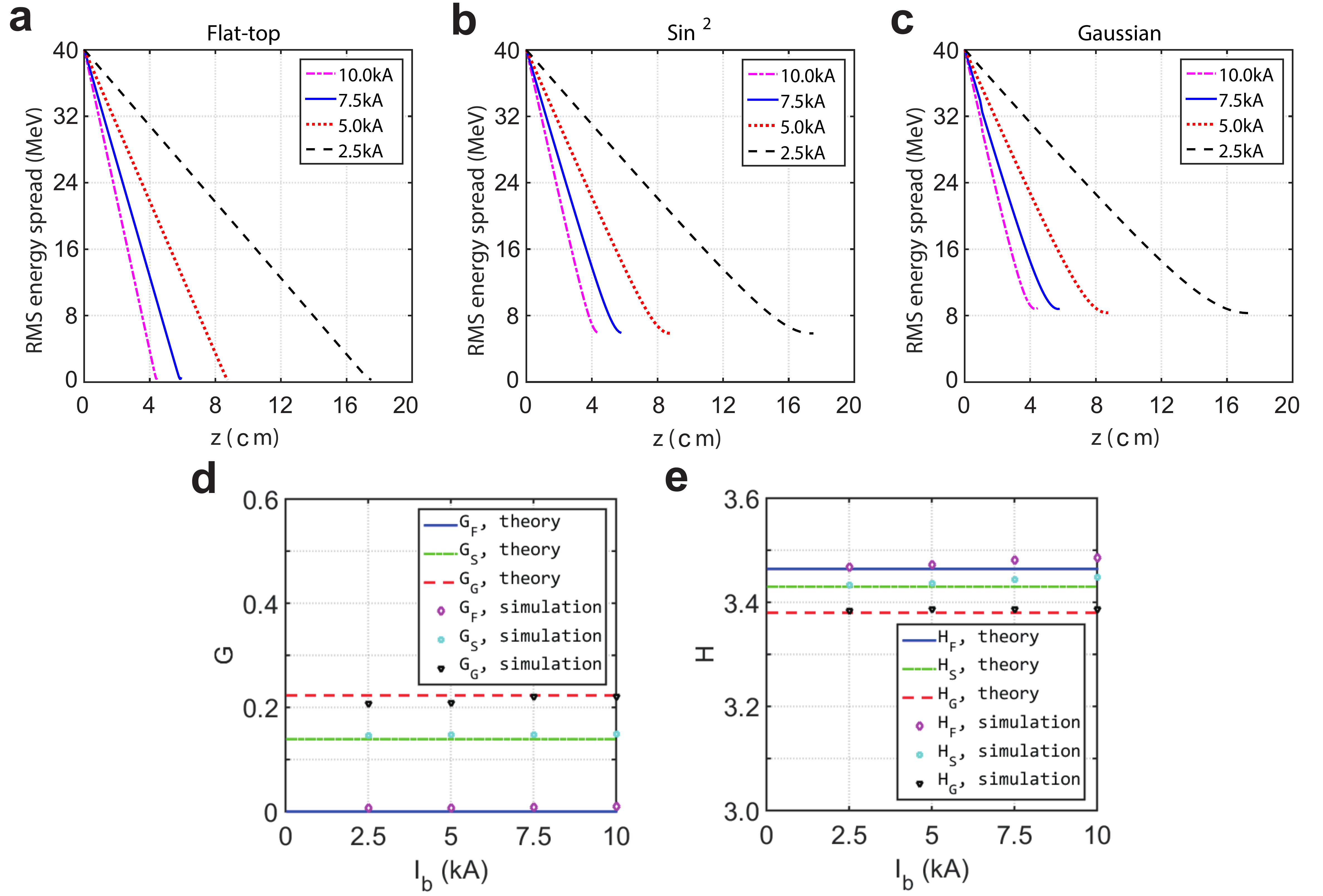}
\caption{\label{fig5} 
The RMS energy spread reduction versus the propagation distances for flat-top (a), sin$^2$ (b) and Gaussian (c) beam profiles with different peak currents.
The other parameters are identical to Fig. \ref{fig3}.
The corresponding calculated and simulated geometrical factors $G$ (d) and $H$ (e) are given for these three current profiles.
}
\end{figure*}

The comparisons of the simulation results with the theoretical expressions are shown in Figs. \ref{fig4} and \ref{fig5}.
Figure \ref{fig4} (a), (b) and (c) show the very good agreement between the calculated and simulated $E_z$ field for three beam profiles (flat-top, sin$^2$ and Gaussian) for a peak current of 10 kA.
For lower beam currents, the agreement will be better due to reduced nonlinear effect.
The corresponding longitudinal phase spaces before and after the dechirper are shown in Fig. \ref{fig4} (d), (e) and (f).
For the flat-top profile, the longitudinal phase space can be completely flattened, leading to a reduction in relative energy spread from $1\%$ to below $0.01\%$.
For the sin$^2$ and Gaussian profiles, $E_z$ has a linear form for most part of the beam except at the bunch head and tail, where nonlinear feature of the wakefield is evident. This nonlinearity results in a sigmoid structure in the longitudinal phase space of the beam.
Despite this non-ideal feature, 
the relative energy spread can still be reduced significantly from $1\%$ to $0.14\%$ (sin$^2$)/$0.22\%$ (Gaussian). 

The energy spread reduction versus propagation distances are also plotted for these three different profiles in Fig. \ref{fig5} (a), (b) and (c). 
It can be clearly seen that the RMS energy spread almost linearly decreases during the beam propagation until the final minimum value is reached. For different peak currents, the achievable minimum RMS energy spreads are almost the same for a given profile, in good accordance with the linear wakefield theory analyses. In Fig. \ref{fig5} (d) and (e), we also plot the calculated and simulated $G$ (that controls the minimal final energy spread $\Delta W_{f}$) and $H$ (that controls the dechirping length $L_c$) factors, and an excellent agreement is achieved.

As one can see, a flat-top beam current profile has the best effect for linear energy chirp reduction, with a capability to obtain energy spread well below $0.1\%$. Indeed, there are several possibilities in plasma-based acceleration for shaping the beam current profile through injection optimizations \cite{PhysRevLett.112.035003, LifeiInjection, Down-ramp_Injection}, and this is a very active research area.

\section{IV. Discussion}

\begin{figure*}[tp]
\includegraphics[height=0.3\textwidth]{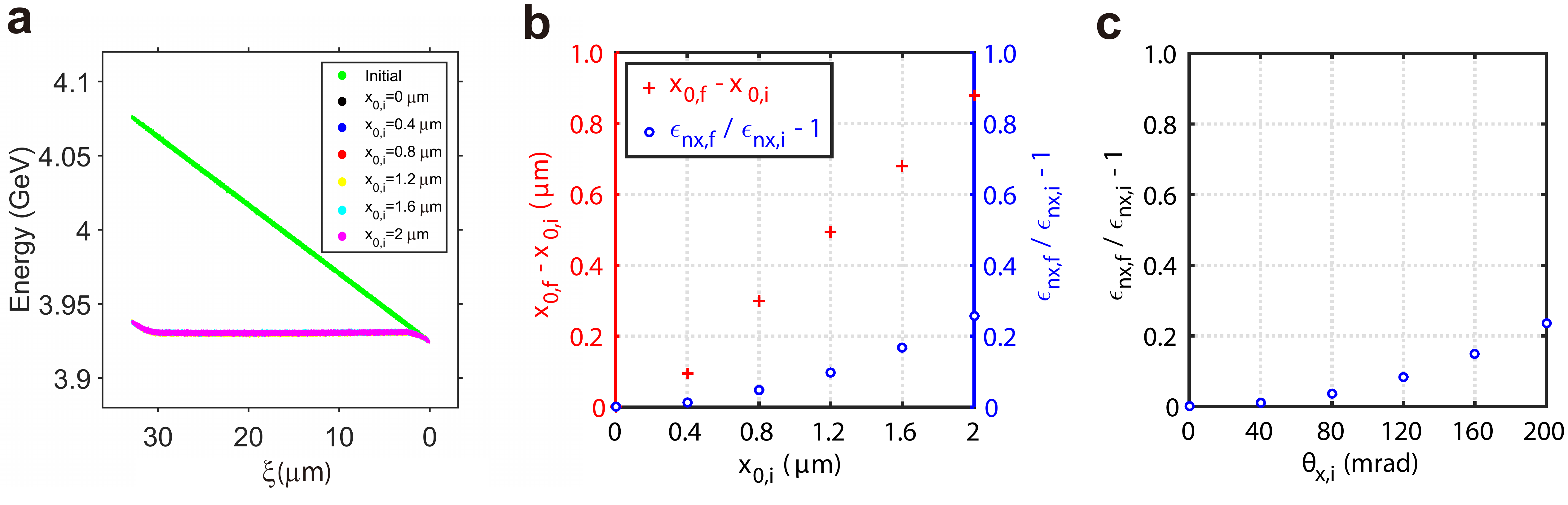}
\caption{\label{fig6}
(a) The initial and final longitudinal phase spaces of the beam for different initial beam offset $x_{0,i}$. Irrespective of the initial offset, the final longitudinal phase spaces are nearly the same. (b) The transverse offset growth of the beam centroid $x_{0,f}-x_{0,i}$ and the projected normalized emittance growth $\epsilon_{nx,f}/\epsilon_{nx,i}-1$ versus the initial beam offset $x_{0,i}$, where the subscripts ``i" and ``f" referring to the initial and final quantities, respectively. (c) The projected normalized emittance growth $\epsilon_{nx,f}/\epsilon_{nx,i}-1$ versus the initial angle of the linear tilt $\theta_{x,i}$. Note that here the beam offset and the tilt are both in $x$ direction.
}
\end{figure*}

The previous sections clearly demonstrate the effectiveness of a hollow channel plasma as a very effective dechirper. In practice, the robustness of this method should also be analyzed. 
A major factor that will derate the performance of this device, is off-axis injection \cite{measurement_of_transverse_wakefields_in_hollow_channel}, which will excite a transverse bending field that will
steer the beam towards one side.
As a result of this deflection, the projected emittance of the beam in the direction of the offset will increase. 
To quantify the tolerance to off-axis injection, the transverse bending field
$W_\perp$ can be calculated for given beam offset and current distribution \cite{thesis_the_positron_driven_hollow_channel_PWFA, measurement_of_transverse_wakefields_in_hollow_channel}, and it turns out that $W_\perp$ is proportional to the offset $x_0$. 
The $W_\perp$ also has a longitudinal dependence of the beam position, thus in the beam transverse phase space,
slice phase-ellipses develop a displacement with respect to each other which leads to the projected emittance growth.
In Fig. \ref{fig6} (a), the final longitudinal phase spaces are plotted for different initial beam offset ($x_{0,i}$) using 3D PIC simulations with beam and plasma parameters identical to Fig. \ref{fig2}.
As expected, small relative offset has a negligible effect on the dechirping effects.
In Fig. \ref{fig6} (b), the transverse beam offset growth and the relative beam projected emittance growth (in the direction of the offset) versus the initial beam offset are plotted. 
As one can see, the beam offset growth is negligibly small compared to the channel inner radius $a$, and the beam emittance growth can be controlled to less than $\sim 25\%$ for initial beam offset within 2 $\mu$m.
In addition to off-axis injection, head-to-tail tilt on an on-axis beam will also lead to transverse wakefields which can induce slice dependent emittance growth. 
In Fig. \ref{fig6} (c), the relative beam projected emittance growth (in the direction of the tilt) versus the initial angle of the linear tilt $\theta_{x,i}$ is shown 
via 3D PIC simulations with the same parameters used in Fig. \ref{fig2}. 
The beam emittance growth can be controlled to less than $\sim 25\%$ for $\theta_{x,i}$ within $\sim 200$ mrad.
These simulation results suggest that the hollow channel plasma dechirper concept has a reasonably good tolerance for both non-ideal off-axis injection and head-to-tail tilt.

\section{V. Summary}
In summary, a new method that uses a hollow channel plasma as a near-ideal dechirper to reduce the energy chirp of electron and positron beams in plasma-based accelerators is proposed. Theoretical analyses and 3D PIC simulations are systematically used to confirm the effectiveness and robustness of this method for reducing the beam energy spread from a few percent level to $\sim 0.1$ percent or lower, while maintaining the beam's slice energy spread and normalized emittance. 
This tunable and flexible technique works well for both electron and positron beams, and it can be applied to future plasma-based photon sources and colliders for significantly enhancing the beam 6D-brightnesses.

\section{ACKNOWLEDGMENTS}
This work is supported by the National Natural Science Foundation of China (NSFC) Grants (No. 11535006, No. 11425521, No. 11775125, and No. 11875175), CAS Center for Excellence in Particle Physics, and the U.S. Department of Energy Grants (No. DE-SC0010064, No. DE-SC0008491, and No. DE- SC0008316) at UCLA.

\section{references}

\section{supplemental material}
\textbf{Calculation of the longitudinal wakefield $E_z$ and geometrical factors ($G$ and $H$) for three typical beam current profiles.}

To quantify the dechirping effectiveness, the evolution of the beam energy spread has been derived for three typical current profiles (flat-top, sin$^2$ and Gaussian).
For an electron/positron beam with a linear positive energy chirp and zero intrinsic slice energy spread, its initial energy profile can be expressed as $W(\xi,z=0)=W_{head,i}+\frac{W_{tail,i}-W_{head,i}}{L_b}\xi$, where $W_{head,i}$/$W_{tail,i}$ is the initial energy of the beam head/tail. 
During the dechirping process, since the decelerating field $E_z$ within the beam has no dependence on $r$, the beam energy distribution at the longitudinal positron $z$ can be written as $W(\xi,z)=W(\xi,z=0)+qE_z(\xi)z$. 
Thus the mean energy and the RMS energy spread of the beam yield
\begin{align}\label{eq5}
W_{mean}(z)=\frac{\int_{0}^{L_b}W (\xi,z)f(\xi)d\xi}{\int_{0}^{L_b}f(\xi)d\xi}
\end{align}
and 
\begin{align}\label{eq6}
\Delta W(z)=\sqrt{\frac{\int_{0}^{L_b}[W(\xi,z)-W_{mean}(z)]^2 f(\xi)d\xi}{\int_{0}^{L_b}f(\xi)d\xi}}
\end{align}
When $z=0$, Eq. \eqref{eq6}  gives the initial RMS energy spread $\Delta W_i$. 
The final minimum RMS energy spread $\Delta W_f$ can be achieved when $\frac{d\Delta W(z)}{dz}=0$ and the corresponding value of $z$ is the required plasma dechirper length $L_c$.

For the flat-top current profile, i.e., $f(\xi)=1$ for $0 \leq \xi \leq L_b$. $E_z$ within the beam is exactly linear along $\xi$
\begin{align}\label{eq7}
E_z(\xi)=-\frac{q}{e}\frac{m k_p^2 c^2}{e}A_0\frac{I_b}{I_A} \xi
\end{align}
Therefore the linear energy chirp of the beam can be completely removed without inducing nonlinear energy chirp or slice energy spread, yielding zero $\Delta W_f$, i.e., $G_F=0$.
The plasma dechirper length $L_{c,F}$ can be intuitively obtained as $L_{c,F}=\frac{W_{tail,i}-W_{head,i}}{|qE_z(\xi=L_b)|}$.
Indeed, to make a comparison between different current profiles, we should keep $\Delta W_i$ fixed. In this case, according to Eq. \eqref{eq6},
$W_{tail,i}-W_{head,i}\approx 3.464 \Delta W_{i}$, 
therefore $L_{c,F}$ can be rewritten in terms of $\Delta W_{i}$ as  
\begin{equation}\label{eq8}
\begin{aligned}
L_{c,F}=\frac{\Delta W_{i}}{|qE_z(\xi=L_b)|}H_F=\frac{\Delta W_{i}}{mk_p^2c^2A_0 \frac{I_b}{I_A}L_b} H_F
\end{aligned}
\end{equation}
where $H_F\approx 3.464$.

For the sin$^2$ current profile, i.e., $f(\xi)=\sin^2(\pi\xi/L_b)$. The expression for $E_z$ is
\begin{align}\label{eq9}
E_z(\xi)=-\frac{q}{e}\frac{m k_p^2 c^2}{e}A_0\frac{I_b}{I_A}\times \left [\frac{\xi}{2}-\frac{L_b\sin(2\pi \xi/L_b)}{4\pi}\right ]
\end{align}
One can see the first term in the brackets is the linear term of $\xi$, while the second term is the high-order nonlinear correction, which leads to the nonlinear energy chirp growth. 
After obtaining the expression for $\Delta W(z)$ and setting $\frac{d\Delta W(z)}{dz}=0$, we find $G_S \approx 0.139$ and $L_{c,S}\approx \frac{W_{tail,i}-W_{head,i}}{|qE_z(\xi=L_b)|}\times 0.6201$. 
Based on Eq. \eqref{eq6}, $W_{tail,i}-W_{head,i}\approx 5.532 \Delta W_{i}$, thus $L_{c,s}$ yields
\begin{align}\label{eq10}
L_{c,S}=\frac{\Delta W_{i}}{|qE_z(\xi=L_b)|}H_S= \frac{\Delta W_{i}}{mk_p^2c^2A_0 \frac{I_b}{I_A}\frac{L_b}{2}}H_S
\end{align}
where $H_S\approx 0.6201\times 5.532  \approx 3.430$.

For the Gaussian current profile, the beam is assumed to be cut off outside three standard deviations from the beam center, i.e., $f(\xi)=\mathrm{e}^{ -\frac{(\xi-L_b/2)^2}{2\times(L_b/6)^2}}$.  $E_z$ within the beam can be expressed as
\begin{align}\label{eq10}
E_z(\xi)=-\frac{q}{e}\frac{m k_p^2 c^2}{e}A_0\frac{I_b}{I_A}\times \sqrt{\frac{\pi}{72}}L_b\left [1+\mathrm{Erf}\left (\frac{\xi-L_b/2}{\sqrt{2}L_b/6} \right ) \right ]
\end{align}
Since the Gauss error function has both the linear and nonlinear terms, high-order nonlinear energy chirp will increase during the dechirping process.
Similar to the sin$^2$ current profile case,  we can obtain 
$G_G \approx 0.223$ and $L_{c,G}\approx \frac{W_{tail,i}-W_{head,i}}{|qE_z(\xi=L_b)|}\times 0.5553$. From Eq. \eqref{eq6} we have 
$W_{tail,i}-W_{head,i}\approx 6.081 \Delta W_{i}$, hence $L_{c,G}$ is given by
 \begin{equation}\label{eq11}
\begin{aligned}
L_{c,G}=\frac{\Delta W_{i}}{|qE_z(L_b)|}H_G= \frac{ \Delta W_{i}}{{mk_p^2c^2A_0\frac{I_b}{I_A}\sqrt{2\pi}L_b/6}}H_G
\end{aligned}
\end{equation}
where $H_G\approx 0.5553 \times 6.081\approx 3.377$.

\end{document}